\documentclass{article}
\usepackage{spconf,amsmath,graphicx,hyperref}
\usepackage{booktabs}
\usepackage{enumitem}
\usepackage{subcaption}
\usepackage{booktabs}
\usepackage{booktabs}   
\usepackage{arydshln}   
\usepackage{makecell}
\usepackage{graphics}
\usepackage{graphicx}      
\usepackage{multirow}
\usepackage{diagbox}

\usepackage{microtype}
\usepackage{graphicx}
\usepackage{arydshln}
\usepackage{inconsolata}
\usepackage{todonotes}
\usepackage{tabularx}
\usepackage{booktabs} 
\usepackage{multirow}
\usepackage{amsmath}
\usepackage{amssymb}
\usepackage{graphicx}
\usepackage{subcaption}
\usepackage{fge}
\usepackage{colortbl}

\usepackage[framemethod=TikZ]{mdframed} 
\usepackage{listings}
\usepackage{longtable}
\usepackage{tcolorbox}
\usepackage{latexsym}
\usepackage{url}
\usepackage{times}
\usepackage{latexsym}
\usepackage{amsmath}
\usepackage{amssymb} 
\usepackage{url}
\usepackage{booktabs}   
\usepackage{arydshln}   
\usepackage{pifont} 
\usepackage{url}

\definecolor{Gray}{gray}{0.92}
\definecolor{racing-green}{rgb}{0.0, 0.8, 0.6}
\definecolor{awesome-red}{rgb}{1.0, 0.13, 0.32}
\definecolor{LightCyan}{rgb}{0.88,1,1}
\definecolor{darkgreen}{RGB}{0,150,0}

\newcolumntype{B}{>{\columncolor{blue!4}}c}
\newcolumntype{d}{>{\columncolor{yellow!4}}c}
\newcolumntype{q}{>{\columncolor{green!4}}c}

\newcolumntype{g}{>{\columncolor{gray!4}}l}

\definecolor{mygray}{gray}{0.95}
\usepackage{arydshln}

\title{Achieving Effective Virtual Reality Interactions via Acoustic Gesture Recognition based on Large Language Models}

\name{Xijie Zhang \qquad Fengliang He \qquad Hong-Ning Dai}
  
\address{Department of Computer Science, Hong Kong Baptist University, Hong Kong, China}
\begin{document}
\ninept
\maketitle
\begin{abstract}
Natural and efficient interaction remains a critical challenge for virtual reality and augmented reality (VR/AR) systems. Vision-based gesture recognition suffers from high computational cost, sensitivity to lighting conditions, and privacy leakage concerns. Acoustic sensing provides an attractive alternative: by emitting inaudible high-frequency signals and capturing their reflections, channel impulse response (CIR) encodes how gestures perturb the acoustic field in a low-cost and user-transparent manner. However, existing CIR-based gesture recognition methods often rely on extensive training of models on large labeled datasets, making them unsuitable for few-shot VR scenarios.
In this work, we propose the first framework that leverages large language models (LLMs) for CIR-based gesture recognition in VR/AR systems. Despite LLMs' strengths, it is non-trivial to achieve few-shot and zero-shot learning of CIR gestures due to their inconspicuous features. To tackle this challenge, we collect differential CIR rather than original CIR data. Moreover, we construct a real-world dataset collected from 10 participants performing 15 gestures across three categories (digits, letters, and shapes), with 10 repetitions each. We then conduct extensive experiments on this dataset using an LLM-adopted classifier. 
Results show that our LLM-based framework achieves accuracy comparable to classical machine learning baselines, while requiring no domain-specific retraining. 
\end{abstract}

%
\begin{keywords}
Virtual Reality, Acoustic Signal, Large Language Model, Gesture Recognition
\end{keywords}
\section{Introduction}
\label{sec:intro}
\begin{figure*}[t]
  \centering
  \includegraphics[width=0.95\textwidth]{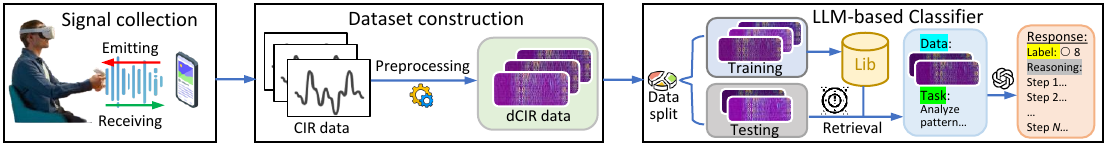}
  \caption{Overview of the proposed framework.}
  \label{fig:experimental setup}
\end{figure*}

Virtual reality and augmented reality (VR/AR) technologies are rapidly expanding beyond entertainment into professional domains such as education, healthcare, and industrial design~\cite{Jiang2022}. A central challenge for VR/AR systems lies in enabling natural, low-latency, and low-computation interactions. Existing gesture recognition modalities face critical limitations.
First, vision-based gesture recognition is computationally expensive, sensitive to lighting conditions, and raises privacy concerns~\cite{CUI2025115693,yang2024can}. 
Second, controller-based input is cumbersome and inefficient~\cite{10379826}. 
Third, voice input is often impractical in quiet public environments~\cite{NUDELMAN20251053}. 
These constraints motivate lightweight, robust, and privacy-preserving alternatives. 

Recently, acoustic sensing has emerged as a promising candidate for gesture recognition in VR/AR systems. In particular, channel impulse response (CIR) offers an attractive means of interaction~\cite{ling2020ultragesture,Wang2022a}. By emitting inaudible high-frequency acoustic signals and capturing the environment’s response, CIR encodes how hand movements perturb the acoustic field. Compared with conventional gesture recognition methods, this approach offers several advantages: it does not rely on visible light, is robust in diverse environments, consumes minimal energy, and remains imperceptible to users, making it especially suitable for resource-limited VR/AR headsets. 

However, current CIR-based recognition methods mostly rely on machine learning (ML) models~\cite{Wang2016,yun2017strata,ling2020ultragesture,Wang2022a}. While effective in constrained settings, these approaches heavily depend on hand-crafted features and exhibit poor generalization, such as $k$-nearest neighbors ($k$NN), random forests (RF), and support vector machines (SVM). Deep learning methods, including convolutional neural networks (CNNs) and pre-trained feature extractors such as CLIP, are powerful but require extensive training on large labeled datasets or domain-specific retraining. Such requirements make them impractical in the few-shot and personalized scenarios in typical VR/AR applications. 

In this work, we explore a new VR/AR gesture recognition approach by leveraging large language models (LLMs) for acoustic gesture recognition. Pre-trained on massive corpora, LLMs exhibit strong few-shot and zero-shot learning capabilities, making them particularly suitable for human activity recognition~\cite{10.1145/3678545}. Compared with traditional ML approaches, LLMs unleash the need for extensive training on CIR data. Moreover, LLMs can infer hand gesture patterns by interpreting underlying gesture dynamics, and they can generalize to unseen gestures under few-shot or even zero-shot conditions. 
Furthermore, LLM-based assistants have been increasingly adopted in VR headsets, making their adoption for hand gesture recognition practically feasible. For instance, the Meta Quest series, one of the most popular VR/AR headsets, now integrates Meta AI~\cite{metaAI-2025}.

However, it is non-trivial to implement such LLM-based gesture recognition. First, there is a lack of a real-world CIR gesture dataset for few-shot and zero-shot learning of LLMs. Second, the CIR raw images contain inconspicuous features, which may not be sufficient for few-shot and zero-shot learning of LLMs. 
To tackle these challenges, we present a framework, which can (1) collect acoustic CIR gesture images, (2) preprocess CIR images into differential CIR (dCIR) images to enhance the CIR features, and (3) instruct an LLM-adopted classifier for hand gesture recognition. The proposed approach is lightweight, low-cost, and privacy-preserved, making it suitable for deployment on VR/AR systems. Our contributions are summarized as follows:
\begin{itemize}
    \item We are the first to explore LLMs for CIR-based gesture recognition. By leveraging inaudible acoustic signals, our method enables robust and privacy-preserving interaction without requiring complex training procedures or hand-crafted features.
    \item We construct a real-world dataset collected from 10 participants performing 15 gestures across three categories (digits, letters, and shapes). 
    \item On a 15-gesture recognition task, we demonstrate that LLMs can classify CIR images in a few-shot setting with accuracy comparable to $k$NN, SVM, and RF baselines, without the need for large-scale training data.
\end{itemize}


\section{Methodology}
\label{sec:methodology}
This section elaborates on the proposed acoustic gesture recognition system. Section~\ref{sec:overall figure}  presents the overview of the framework. Section~\ref{sec:data collection} then illustrates the steps of preprocessing acoustic signals. Section~\ref{sec:core function} next introduces the core design of gesture recognition via LLMs, including image retrieval and multimodal prompt building.

\subsection{Overview of the main method}
\label{sec:overall figure}

Fig.~\ref{fig:experimental setup} briefly illustrates the framework for acoustic gesture recognition. In particular, a participant sits at a desk, wearing a VR headset and holding a controller to perform diverse operations (e.g., handwriting). A smartphone is placed nearby, continuously emitting and receiving inaudible acoustic signals. After collecting and preprocessing the received acoustic signals, we then obtain a series of dCIR images, which were subsequently converted into a dCIR image set. Next, the image set is divided into a training set and a testing set, which are used for the library construction, retriever, and feeding into prompts for LLMs. Finally, we obtain the response, i.e., the classification of handwriting types with reasoning steps.

\subsection{Preprocessing acoustic signals}
\label{sec:data collection}

Emitted from a mobile phone nearby, a high-frequency ($>$ 20 kHz) acoustic signal experiences reflections, diffusions, and diffractions when passing through the participant's hands, which move to form hand gestures. From the received signals, we then obtain the channel impulse response (CIR) patterns of each gesture. After preprocessing the CIR raw data, we then obtain differential CIR (dCIR) images.

\textbf{Signal Generation and Collection.}
A 50-bit baseband frame was constructed from a 26-bit GSM training sequence due to its efficiency in channel estimation and synchronization. The sequence was up-sampled by 12× replication interpolation, smoothed with a 2 kHz low-pass filter, and up-converted to a 20 kHz center frequency in the ultrasonic band. After band-pass filtering, the signal was transmitted through the smartphone speaker, and reflections were eventually recorded by the microphone of the mobile phone.

\textbf{Signal Processing.}
The received signal was down-converted and low-pass filtered to suppress environmental noise. Frame boundaries were detected via the Pearson Correlation Coefficients with the original sequence, and each frame was segmented and min–max normalized. Subsequently, CIR denoted by $\chi(t)$ was then estimated using the Least Squares method, and its differential form (aka dCIR) $\delta(t) = \chi(t) - \chi(t-1)$ was computed to capture the acoustic channel variations caused by hand movements.
Profiling the corresponding gesture, the resulting dCIR image was used as the input for subsequent LLM-based recognition.




\subsection{Gesture recognition via LLMs}
\label{sec:core function}
\begin{figure}[h]
    \centering
    \includegraphics[width=0.95\linewidth]{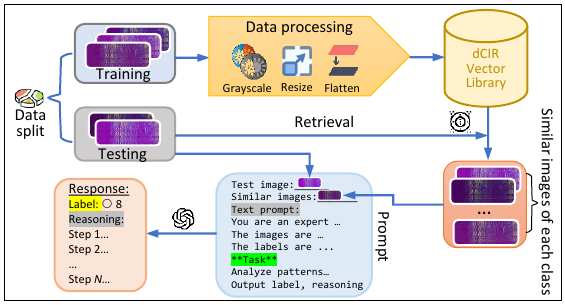}
    \caption{LLM-based Classifier for Gesture Recognition}
    \label{fig:main function}
\end{figure}

Fig.~\ref{fig:main function} shows a detailed workflow of the core gesture recognition module. Since directly appending all training images to the input of an LLM is computationally infeasible and often unsupported by context length limitations, a lightweight dataset and an efficient retrieval mechanism are essential to identify a small set of relevant reference images for few-shot gesture recognition. To achieve this goal, during image pre-processing, each image in the training set was converted to grayscale, reducing the computation cost. Next, it was scaled to a uniform size of 64×64 pixels and flattened to a 4096-dimensional feature vector to accommodate subsequent model processing. All processed vectors of each gesture class were compiled into the corresponding dCIR vector library. 
For a test image, the same preprocessing pipeline was applied to extract its feature vector. A $k$NN model was then used to retrieve the most similar vectors (i.e., dCIR images) from the vector library of each class, and each vector corresponds to a dCIR image, its gesture label, and an Euclidean distance to the test vector. Subsequently, the test image, the retrieved samples, their associated gesture labels, and distances were integrated into a multimodal (image + text) prompt and submitted to the LLM for gesture recognition. The LLM's output contained its determined gesture label and the corresponding reasoning.

\begin{figure}[h]
    \centering
    \includegraphics[width=0.95\linewidth]{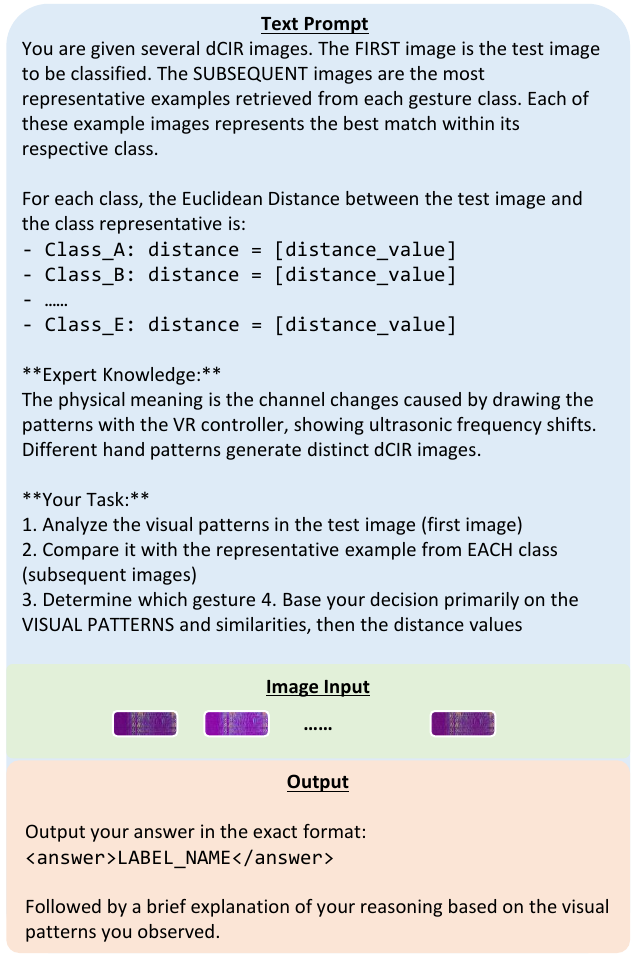}
    \caption{An example visual prompt for LLMs}
    \label{fig:prompt}
\end{figure}

Fig.~\ref{fig:prompt} presents a simplified example prompt for vision-driven gesture recognition. As mentioned above, the input consists of: (1) a test dCIR image, and (2) the most similar samples of each class retrieved from the training database. Each sample is annotated with its gesture label and the corresponding Euclidean distance to the test image.
The system message initializes the LLM as an expert in dCIR image analysis. The user message then presents some basic knowledge about the source of the images, followed by each sample's label and Euclidean distance. The physical meaning of the labels is also provided, namely, channel changes caused by drawing the label patterns with the VR controller.
The core instruction, starting with \texttt{\small **Your Task**}, requires the LLM to analyze the patterns in the dCIR images to determine which provided gesture pattern the test image most closely matches. Especially, the LLM is instructed to perform visual reasoning as a major task, rather than focusing on the distance analysis. The expected output is structured as: $<$\texttt{\small answer}$>$\texttt{\small LABEL}$<$\texttt{\small /answer}$>$, followed by a step-by-step reasoning on how the visual patterns were compared and interpreted.

\section{Experiments}
\label{sec:experiments}
This section presents the experimental evaluation of the proposed gesture recognition system. Section~\ref{sec:dataset} describes the constructed dataset used in the experiments. Section~\ref{sec:evaluation protocol} outlines baselines and performance metrics adopted for evaluation. Section~\ref{sec:param} details the key parameters of the system components. Finally, Section~\ref{sec:result} presents experimental results.


\subsection{Experimental Setup} \label{sec:dataset}
\begin{figure}[t]
    \centering
    \includegraphics[width=\linewidth]{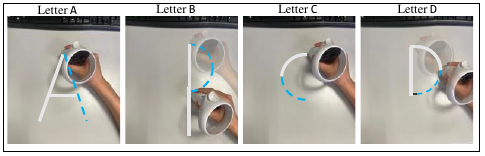}
    \caption{Handwriting letters by VR controllers}
    \label{fig:ex-letters}
\end{figure}

We recruited 10 participants to perform hand gestures while wearing a VR headset and holding VR controllers. We considered three gesture categories: (i) \textit{digits}: 1–5, (ii) \textit{letters}: A–E, and (iii) \textit{shapes}: circle ($\bigcirc$), diamond ($\Diamond$), triangle ($\bigtriangleup$), check mark ($\checkmark$), and cross ($\times$). In total, this yielded 15 distinct gesture classes. Each participant repeated every gesture 10 times, resulting in a dataset of $10 \times 15 \times 10 = 1500$ CIR samples. The emitted acoustic signals were inaudible high-frequency tones, ensuring that the sensing was unobtrusive to participants. Fig.~\ref{fig:ex-letters} shows examples of handwriting letters in the experiment. This study is approved by our university’s Institutional Review Board (IRB). All participants are informed of the experiment and compensated for their time.

For each trial, the device continuously emitted the acoustic signal and recorded the environmental reflections. CIR sequences were extracted from the received signals and converted into two-dimensional images for classification. To minimize inter-participant variability, all data were collected under the same environmental conditions. Each CIR image was normalized and resized to a fixed resolution before being provided to the models. Upon data collection, the complete dataset was partitioned for each gesture class into training and testing sets, using an 80/20 ratio.
\subsection{Evaluation Baselines and Metrics} 
\label{sec:evaluation protocol}
We evaluated the proposed LLM-based framework against three classical machine learning baselines: $k$NN, RF, and SVM. For fair comparison, all models received the same CIR image inputs. In the LLM setting, we adopted a few-shot prompting strategy, where a small number of labeled CIR samples were provided as in-context examples, while the remainder were used for testing. For the baseline models, standard 80/20 train-test splits were applied. The retrieval and LLM calling process were applied to all test samples, strictly following the workflow of the core module. After collecting the feedback from the LLMs, error responses were removed. Then, the segments enclosed within the $<$\texttt{\small answer}$>$ and $<$/\texttt{\small answer}$>$ tags are extracted from each remaining response, representing the classification results generated by the LLM. The rest of the response, indicating LLM’s reasoning, was left alone for reference.

Performance was assessed using classification precision $\eta$, recall ($\rho$), and F1-score (F1) across the 15 gesture classes. We further reported the accuracy of each category (digits, letters, shapes) to analyze differences in gesture complexity. In addition to accuracy, we qualitatively evaluated the interpretability of LLM predictions by examining generated explanations for unseen CIR inputs in a zero-shot setting.

\begin{table*}[t]
\centering
\resizebox{\textwidth}{!}{
\scriptsize
\begin{tabular}{l|ccc|ccc|ccc}
\toprule
\multirow{2}{*}{\diagbox{Models}{Categories}} & 
\multicolumn{3}{c|}{\text{Shapes} ($\bigcirc$$\Diamond$$\bigtriangleup$$\checkmark$$\times$)} &
\multicolumn{3}{c|}{\text{Letters} (\texttt{A}, \texttt{B}, \texttt{C}, \texttt{D}, \texttt{E})} &
\multicolumn{3}{c}{\text{Digits} (\texttt{1}, \texttt{2}, \texttt{3}, \texttt{4}, \texttt{5})} \\
 & Precision ($\eta$) & Recall ($\rho$) & F1 & Precision ($\eta$) & Recall ($\rho$) & F1 & Precision ($\eta$) & Recall ($\rho$) & F1 \\
\midrule
Claude 4 Sonnet   & 0.901 & 0.896 & 0.896 & 0.890 & 0.888 & 0.888 & 0.912 & 0.906 & 0.906 \\
Gemini 2.5 Pro    & 0.884 & 0.878 & 0.878 & 0.826 & 0.786 & 0.777 & 0.897 & 0.877 & 0.879 \\
GPT-5             & 0.897 & 0.890 & 0.889 & 0.861 & 0.861 & 0.861 & 0.936 & 0.935 & 0.935 \\
\midrule 
SVM               & 0.929 & 0.927 & 0.927 & 0.902 & 0.882 & 0.883 & 0.932 & 0.930 & 0.930 \\
$k$NN               & 0.933 & 0.933 & 0.933 & 0.917 & 0.904 & 0.905 & 0.946 & 0.942 & 0.942 \\
RF & 0.858 & 0.852 & 0.852 & 0.963 & 0.765 & 0.852 & 0.919 & 0.918 & 0.917 \\
\bottomrule
\end{tabular}
}
\caption{
Weighted average performance of various models across three gesture categories: shapes, letters, and digits.
}
\label{tab:performance table}
\vspace{-0.4cm}
\end{table*}

\subsection{Parameters and Settings}
\label{sec:param}
We conducted the experiments based on the three gesture categories: shapes, letters, and digits, where each category contains five gesture classes. For each trial, the prompt was appended with one test image and five retrieved sample images from the same category. This approach enhances the LLM's performance by providing contextually similar patterns while maintaining a manageable computational load.
Major powerful multimodal LLMs, including Gemini 2.5 Pro~\cite{GoogleDeepMind2025}, ChatGPT-5~\cite{OpenAI2025} and Claude 4 Sonnet~\cite{Anthropic2025}, were evaluated using their APIs. The maximum tokens were set to 4096 for detailed reasoning output, and the temperature was set to 0.2 for more consistent responses. Other parameters were set to the default values.

\subsection{Experimental results}
\label{sec:result}

Table~\ref{tab:performance table} presents per-class precision ($\eta$), recall ($\rho$), and F1-score across three LLMs, as well as baseline ML methods. 
As shown in Table~\ref{tab:performance table}, all three LLMs achieved satisfactory performance in recognizing correct gestures, with most F1-scores exceeding 0.8. This indicates that not only the most similar images are accurately retrieved, but the LLMs also possess strong capabilities in processing multi-modal inputs, dCIR patterns, and matching them to the correct labels.

Overall, the LLMs demonstrated more stable and accurate classification performance for the ``Shapes'' and ``Digits'' categories, with all corresponding metrics exceeding 0.85. This excellent performance can be attributed to the distinct kinematic characteristics involved in drawing shapes and digits. Most shapes are composed of straight lines with clear directional changes and vertices. Similarly, digit `\texttt{1}' through digit `\texttt{3}' can typically be drawn in a single stroke, and digit `\texttt{4}' and digit `\texttt{5}'s writing are highly distinct. As a result, these gestures produce simpler, more structured, and unique dCIR patterns, characterized by hand and arm movements. Such patterns appear to be more interpretable and distinguishable for LLMs.

On the contrary, writing letters requires more curves and uncertain turning points, resulting in more fluctuated classification results. Only letter `\texttt{C}' requires a single stroke, while writing letter `\texttt{B}' requires more changes in gesture direction and starting points between strokes, making its patterns more easily confused with those of other letters, for example, letter `\texttt{D}'. As a result, their precisions were relatively low across all three LLMs, significantly lowering the overall performance of the ``Letters'' category.

Meanwhile, three LLMs performed differently on each category.  GPT outperformed the other two LLMs on recognizing the dCIR patterns of digits, while Claude better distinguishes the dCIR images of letters. However, there is still a performance gap between LLM and traditional ML methods. In Digits classification, GPT narrows the gap to just 0.007 compared to $k$NN, but in the more complex Letters category, even the best-performing LLM trails the most accurate ML method by 0.017, highlighting the current limitations of LLMs for sophisticated dCIR pattern recognition tasks. We also examined LLMs' performance on zero-shot learning by manually checking the response, and the result is not satisfactory enough. This is probably attributed to the visual encoding mechanisms or pre-training data. For general-purpose design LLMs, there is relatively less training for professional and specialized data graphs like dCIR images. 

Notably, the most significant advantage of LLMs lies in their capacity to generate reasoning, thereby enhancing the interpretability for human users, which is nevertheless lacking in ML methods. Moreover, LLMs are undergoing rapid advancement, steadily closing the performance gap with specialized ML approaches. Through fine-tuning, LLMs are not only capable of tackling specialized tasks but also hold promising potential for broader applications. 
\section{Related Work}
\label{sec:related}

Gesture recognition has been a crucial challenge in VR/AR interactions. Commercial off-the-shelf (COTS) VR devices are dominated by vision-based approaches~\cite{Oyedotun2017, Sharma2021, Pan2024}, which often incur high computational costs. These limitations motivate the exploration of non-visual and lightweight modalities.  

\subsection{Acoustic Sensing and CIR-based Methods}
Acoustic sensing has recently emerged as a promising alternative for gesture recognition~\cite{deshotels2014inaudible} and other audio applications~\cite{10446838,10446507}. CIR-based approaches have demonstrated outstanding performance in gesture recognition~\cite{ling2020ultragesture} with the ability to classify simple gestures such as shapes and strokes~\cite{YWang:TMC22, yun2017strata,chen2022swipepass}. Nevertheless, most existing methods rely heavily on ML models that require large labeled datasets, which is impractical in few-shot VR scenarios. Others mainly target mobile interactions and overlook the large-scale hand movements characteristic of VR environments. Moreover, current approaches are unable to explain the relationship between CIR patterns and gestures, thereby limiting their extensibility. In contrast, our LLM-based recognition framework achieves high accuracy under few-shot conditions while also offering interpretability and zero-shot generalization.

\subsection{LLMs for Multimodal Reasoning}
LLMs have recently shown strong capabilities in interpreting structured and multimodal data beyond text. For sensor-based inputs, LLMs have been applied to IMU streams, physiological signals, and even raw audio, where simple visualizations or embeddings enable few-shot or zero-shot reasoning~\cite{Yan2024, Yoon2024, Sheikh2024, qu2024llms, Liang2025}. For audio data, techniques such as acoustic prompt tuning can generate contextual embeddings, enabling LLMs to perform reasoning over auditory tasks~\cite{Liang2025}. However, to the best of our knowledge, no prior work has explored the use of LLMs for gesture recognition from CIR images.

\section{Conclusion}
\label{sec:conclusion}
This paper presents an LLM-based hand gesture recognition system in VR environments. We collect dCIR acoustic data via real-world experiments by 10 participants who perform 15 gestures by holding VR controllers. After data augmentation, the dCIR data is then subsequently partitioned to form a lightweight dCIR vector library. We then design a multimodal prompt with integration of text instruction, a test image, and retrieved images for LLMs. Experimental results show that the system achieves satisfactory accuracy in classifying various gestures, thus providing a reliable approach for understanding interactions in VR environments. Future work includes an in-depth analysis of the reasoning process by LLMs during gesture recognition. Moreover, we will also explore fine-tuning and other techniques to fully unleash the potential of LLMs in VR/AR systems.

\vfill\pagebreak




\bibliographystyle{IEEEbib}
\bibliography{refs}

\end{document}